\documentstyle[aps,manuscript]{revtex}

\begin{document}

\title{The effect of Cu-doping on the magnetic and transport properties of 
La$_{0.7}$Sr$_{0.3}$MnO$_{3}$}

\author{M. S. Kim$^{1,2}$, J. B. Yang$^{1,2}$, Q. Cai$^4$, X. D. Zhou$^2$, W.
J. James$^{2,3}$, W. B. Yelon$^{2,3}$,  P. E. Parris$^1$, D. Buddhikot$^5$, S. 
K. Malik$^5$}

\address{$^1$Department of Physics, University of Missouri-Rolla, Rolla,
MO 65409}
\address{$^2$Graduate Center for Materials Research, University of
Missouri-Rolla, Rolla, MO 65409}
\address{$^3$Department of Chemistry, University of Missouri-Rolla, Rolla,
MO 65409}
\address{$^4$Department of Physics, University of Missouri-Columbia, Columbia,
MO 65211}

\address{$^5$Tata Institute of Fundamental Research, Colaba, Mumbai 400-005,
India}

\maketitle

\begin{abstract}
The effects of Cu-doping on the structural, magnetic, and transport properties 
of La$_{0.7}$Sr$_{0.3}$Mn$_{1-x}$Cu$_{x}$O$_{3}$ (0 $\leq$ x $\leq$ 0.20) have 
been studied using neutron diffraction, magnetization and magnetoresistance(MR) 
measurements. 
All samples show the rhombohedral structure with the \emph{R$\overline{3}$c} 
space-group from 10K to room temperature(RT). Neutron diffraction data suggest 
that some of the Cu ions have a Cu$^{3+}$ state in these compounds. 
The substitution of Mn by Cu affects the Mn-O bond length and Mn-O-Mn bond 
angle resulting from the minimization of the distortion of the MnO$_{6}$ 
octahedron. Resistivity measurements show 
that a metal to insulator transition occurs for the x $\geq$ 0.15 samples. The 
x = 0.15 sample shows the highest MR($\approx$80$\%$), which might result from 
the co-existence of Cu$^{3+}$/Cu$^{2+}$ and the dilution effect of Cu-doping on 
the double exchange interaction.


\end{abstract}

\section{Introduction}

The ABO$_{3}$ (A=trivalent rare earth atom, B=divalent alkaline metal) 
perovskites  have attracted considerable attention because of the anomalous 
magnetic and transport properties such as colossal magnetoresistance(CMR), 
metal-insulator transitions(MIT).\cite{goodenough,unishibara,tomioka} The 
La$_{1-x}$Sr$_x$MnO$_3$ pervoskites are a canonical ABO$_{3}$-type, CMR 
material.  It has been reported that  double-exchange(DE) interactions together 
with Jahn-Teller(JT) distortions in the MnO$_{6}$ octahedron may be required to 
account for the large MR effects.\cite{zener,anderson,kubo,millis}
Because the Mn ion is the center of the DE interaction, the role of the Mn atom 
and its local environment have become the focus of much of the research on 
manganese perovskites. To investigate the CMR associated with the lattice 
deformation and  charge ordering of the crucial Mn-O-Mn network, many early 
studies were carried out by doping the A-site with divalent atoms(Ca, Sr, Ba, 
etc).\cite{neumeier,medarde,cao}  Recently, it has also been shown that 
substitution of Mn(B-site) by other atoms dramatically affects the magnetic and 
transport properties of manganese perovskites.\cite{martin,blasco,ahn,yuan} The 
B site modification directly affects the Mn network by changing the 
Mn$^{3+}$/Mn$^{4+}$ ratio and the electron carrier density, which may provide a 
better understanding of the mechanism for CMR effects in the perovskites.  

In this study, we report the effects of replacing Mn with Cu on the structural, 
magnetic and transport properties of  
La$_{0.7}$Sr$_{0.3}$Mn$_{1-x}$Cu$_{x}$O$_{3}$ with 0 $\leq$ x $\leq$ 0.20. 

\section{Experimental}
Samples of Cu-substituted La$_{0.7}$Sr$_{0.3}$Mn$_{1-x}$Cu$_{x}$O$_{3}$, with 
0 $\leq$ x $\leq$ 0.20, were prepared using the conventional solid state 
reaction 
method. Highly purified La$_{2}$O$_{3}$, SrCO$_{3}$, CuO, MnO were mixed 
in stoichiometric ratios, ground, and then pelletized under 3,000 psi pressure 
to a 1 cm diameter. The pelletized samples were fired at $1350^{\circ}C$ in air 
for 4 1/2 hours, then reground and sintered at $1100^{\circ}C$ for 24 hours. 
Sintered samples were cooled to $900^{\circ}C$ in the oven and removed from the 
furnace, then finally cooled to RT in air. X-ray diffraction of the powders was 
carried out at RT using a SCINTAG diffractometer with Cu-K$\alpha$ radiation. 
Powder neutron diffraction experiments were performed at the University of 
Missouri-Columbia Research Reactor(MURR) using neutrons of wavelength 
$\lambda$ = 1.4875 $\AA$. Refinement of the neutron diffraction data was 
carried 
out using the FULLPROF program.\cite{full} Magnetic measurements were conducted 
with a SQUID magnetometer and resistivity data were obtained using a physical 
properties measurement system(PPMS) with a standard four-point probe method.

\section{Results and Discussion}

X-ray diffraction studies of La$_{0.7}$Sr$_{0.3}$Mn$_{1-x}$Cu$_{x}$O$_{3}$ 
samples with 0 $\leq$ x $\leq$ 0.20 indicate that all samples are single phase 
and all peak positions can be indexed to 
La$_{0.67}$Sr$_{0.33}$MnO$_{2.91}$(JCPDS 50-0308). To further investigate the 
structure distortion and magnetic properties of these compounds, powder neutron 
diffraction data were collected at RT and 10K. The typical neutron diffraction 
patterns of La$_{0.7}$Sr$_{0.3}$Mn$_{0.9}$Cu$_{0.1}$O$_{3}$ measured at 10K and 
RT are shown in Fig. 1. All patterns can be fitted well with the 
$R$$\overline{3}$c rhombohedral space-group (No.167). Refined lattice 
parameters and magnetic moments for the compounds are listed in Table I. At RT, 
the lattice parameters $a$,$c$ and the unit cell volumes decrease with 
increasing Cu content. At 10K, the lattice parameters and unit cell volume 
remain nearly constant with increasing Cu content. The refined magnetic moments 
of Mn atoms at 10 K and RT decrease with increasing Cu content. These results 
agree well with the values obtained from magnetic measurements.

Generally, the B-site doping will directly change the Mn$^{3+}$/Mn$^{4+}$ ratio 
and the exchange interaction of Mn-Mn. The lattice parameters and crystal 
structure will be affected due to the mismatch of the ionic radius between the 
Mn ions and the doping ions. Even though the most stable state of Cu is 
Cu$^{2+}$, the Cu$^{2+}$(6 coordination) radius is about 0.73$\AA$,  which is 
much larger than the radius of Mn$^{3+}$ (0.645 \AA) and Mn$^{4+}$ 
(0.53\AA).\cite{shannon} The substitution of Mn by Cu$^{2+}$ will lead to an 
expansion of the unit cell rather than a contraction. 
Therefore, the decrease of the unit cell volume with Cu-doping at RT suggests 
that some of the Cu ions are in a Cu$^{3+}$ state with a radius of 0.54 \AA, 
which is smaller than that of Mn$^{3+}$ and larger than that of Mn$^{4+}$. 
Furthermore, this is able to account for the valence states of Mn in order to 
satisfy the charge balance in these compounds. It should be pointed out that 
both Cu$^{2+}$ and Cu$^{3+}$ states may appear in these compounds, which might 
be the reason that the changes of the unit cell volume and the lattice 
parameters are not linear with Cu content. A similar phenomenon has been 
observed in (La,Ba)Cu$_{1-x}$Mn$_x$O$_3$ compounds, where Cu$^{3+}$ 
enters Mn sites. \cite{yuan}

The average (Mn,Cu)-O bond length and (Mn,Cu)-O-(Mn,Cu) bond angle extracted 
from the Rietveld refinements are shown in Fig. 2.  At RT, the average bond 
length decreases gradually while the average bond angle increases up to 
x = 0.10 then slightly decreases for $x \geq$ 0.15. At 10K, the average bond 
length decreases up to x = 0.1 and remains constant for x $\geq$ 0.15, while 
the 
average bond angle increases and attains a maximum value at x = 0.10. The 
average bond length and the average bond angle are directly related through the 
oxygen positions and content. The refinement results indicate that there is no 
oxygen deficiency within error limits. Therefore the unusual changes in the 
average (Mn,Cu)-O bond length and (Mn,Cu)-O-(Mn,Cu) bond angle between RT and 
10K might be related to the magnetostriction that reflects the magnetic 
ordering temperature and the existence of mixed Cu$^{2+}$/Cu$^{3+}$ states. The 
internal strain in the MnO$_{6}$ octahedron may be released by a small 
distortion of the MnO$_{6}$ octahedra through 
changes in bond length and bond angle. 

Fig. 3 shows the magnetization versus temperature (M-T) curves measured under 
field cooling (FC) and zero field cooling (ZFC) conditions in a magnetic field 
of 50 Oe for the x = 0.05, 0.10, and 0.15 samples. Two magnetic transition 
temperatures are shown for the samples with x = 0.05 and 0.10. 
The neutron diffraction data do not show the presence of other phases in the 
Cu-doped samples within the resolution of neutron diffraction analysis. 

The electronic bandwidth $W$ has been used to explain the changes in the 
magnetic transition temperatures by varied A and B-site doping.\cite{radaelli} 
The empirical formula of the bandwidth $W$ for ABO$_{3}$-type perovskites using 
the tight binding approximation\cite{harrison} is $W \propto 
\frac{cos\omega}{(d_{Mn-O})^{3.5}}$, where $\omega = \frac{1}{2} (\pi 
-\theta_{\langle Mn-O-Mn \rangle})$, $d_{Mn-O}$ is the average (Mn,Cu)-O bond 
length, and $\theta_{\langle Mn-O-Mn\rangle}$ is the average (Mn,Cu)-O-(Mn,Cu) 
bond angle. The calculated relative bandwidth $W'$ is listed in Table I. The 
bandwidth at RT is smaller than that at 10K for a given Cu content. The change 
across the series is very small. At RT, the increase in the bond angle and the 
decrease in bond length lead to an increase of the bandwidth which increases 
the overlap between the O-$2p$ and the Mn-$3d$ orbitals. Therefore, the 
Cu-doping could enhance the exchange coupling of Mn$^{3+}$-Mn$^{4+}$, and 
increase the magnetic ordering temperature T$_C$ at a low Cu-doping ratio. This 
is consistent with the fact that the low Cu-doped samples ($x \leq$ 0.05) show 
almost no decrease in Curie temperature T$_{C}$.  At a high doping ratio, the 
magnetic dilution effect of Cu is predominant, which gives rise to a sharp drop 
in the Curie temperature for $x>0.10$.
 
Fig. 4 shows the temperature dependence of resistivity under various applied 
fields for La$_{0.7}$Sr$_{0.3}$Mn$_{1-x}$Cu$_{x}$O$_{3}$ ($x$ 
=0.10, 0.15, and 0.20). With increasing Cu content, the resistivity of the 
compound increases, while the resistivity decreases with increasing magnetic 
field. This is ascribed to a reduction of the Mn$^{3+}$/Mn$^{4+}$ ratio to 
account for the DE interaction and a reduction in the number of hopping 
electrons and hopping sites by Cu substitution. In addition, Cu$^{3+}$ has 8 
electrons in the $d$-orbital, and is highly localized with a strong 
Coulomb repulsion. Therefore, the Mn$^{3+}$ $e_{g\uparrow}^{1}$ electron may 
not itinerate through the Mn$^{3+}$-Cu$^{3+}$ chains but through the 
antiferromagnetic super-exchange chains of  Mn$^{3+}$-Mn$^{3+}$ and 
Mn$^{4+}$-Mn$^{4+}$. 
Therefore Cu substitution weakens the DE interaction, disturbs the Mn-O-Mn 
network, and creates short range ordered ferromagnetic clusters. As more Cu is 
substituted, more inhomogeneous small clusters are formed, leading to a 
broadening of the magnetic phase transition peak. Consequently, Cu-doping 
enhances antiferromagnetic super-exchange interactions and weakens DE 
interactions which reduces electron hopping sites and increases electronic 
resistivity.  

For the $x \leq 0.10$ samples, metallic behavior is shown with decreasing 
temperature and for the samples with $x \geq 0.15$, a metal-insulator 
transition 
(MIT) appears.(see Fig. 3) A resistivity peak corresponding to the magnetic 
transition is present. There is no clear field-induced shift of 
maximum resistivity for all samples.  The suppression of the resistivity by the 
applied magnetic field occurs over the entire temperature range for all 
samples. In the DE mechanism, the mobility of the charge carrier $e_{g}$ 
electrons improves if the localized spins are polarized. The applied field 
aligns the canted electron spins which should reduce the scattering of 
itinerant electrons with spins and thus the resistivity is reduced. The 
temperature dependence of the magnetoresistance was calculated with [MR = 
$(\rho_{0}-\rho_{H}) / \rho_{0} \times 100$] under H=1, 3 and 5T. The MR 
increases with increasing Cu content up to $x \leq 0.15$ samples and shows the 
maximum MR=80$\%$  below T=100K, for the $x=0.15$ sample. The maximum MR 
decreases with further Cu-doping. The bandwidth, $W$, which is associated with 
the small structure distortion cannot alone account for the CMR behavior of 
this system.  The existence of Cu$^{3+}$/Cu$^{2+}$, the dilution effect of Cu 
on the DE interaction and the 
super-exchange interaction of Mn may be responsible for the MR behavior in 
these compounds.

In summary, we have investigated the structural, magnetic and electronic 
properties of Cu-doped La$_{0.7}$Sr$_{0.3}$Mn$_{1-x}$Cu$_{x}$O$_{3}$. All 
samples show the same crystal structure from 10 K to RT. The variations of the 
bond length and bond angle resulting from
Cu substitution minimize the distortion of the MnO$_6$ network and stabilize 
the rhombohedral structure. A mixture of Cu$^{2+}$ and Cu$^{3+}$ ions  gives 
rise to a decrease of the unit cell volume and a change of the Mn valence 
states in these compounds. The metal to insulator transition for the samples 
with x $\geq 0.15$ result from changes in the Mn-O-Mn interaction, 
bandwidth, Mn valence state and the charge carrier concentration in the 
Cu-doped compounds. 

\section{ACKNOWLEDGMENTS}
We thank Aranwela Hemantha for invaluable help with the  magnetoresistance
measurements. The support by DOE under DOE contract \#DE-FC26-99FT400054 is
acknowledged. 

\newpage
\baselineskip = .5\baselineskip

\small
\begin{table}{Table I. Refined parameters for 
La$_{0.7}$Sr$_{0.3}$Mn$_{1-x}$Cu$_{x}$O$_{3}$, \emph{R$\overline{3}$c} 
space-group, at room temperature and T=10K. Numbers in parentheses are 
statistical errors. $a$ and $c$ are the lattice parameters. $m$ is magnetic 
moment. $V$ is the unit cell volume. $\chi^2$ is [R$_{wp}$/R$_{exp}$]$^2$ where 
R$_{wp}$ is the residual error of the weighted profile. The magnetic moments of 
the x $\geq 0.15$ samples at RT are not refined. Electronic bandwidth 
parameter $W'$=10*$cos\omega/(d_{Mn-O})^{3.5}$(arb. unit).}

\begin{center}
\begin{tabular}{ccccccc}
Composition (x)   & 0.00          & 0.05        & 0.10        & 0.15        & 
0.20 \\ \hline
&& T=300K&&& \\ \hline
$a$ ($\AA$)     & 5.5038(2)   & 5.5003(2)   & 5.4987(2)   & 5.4982(2)   & 
5.4941(2)\\
$c$ ($\AA$)     & 13.3552(5)  & 13.3387(4)  & 13.3341(5)  & 13.3304(6)  
&13.3180(6)\\
$V$ ($\AA^{3}$) & 350.351(18) & 349.473(17) & 349.153(18) & 348.990(22) 
&348.152(24)\\
$m$ ($\mu_{B}$) & 2.512(28)   & 1.975(30)   & 1.411(101)   & -  
& -  \\
$\chi^{2}$ ($\%$) & 2.81      & 2.69        & 2.91        & 4.89        & 5.36 
\\ 
$W'$ 	        &0.953        & 0.956       &0.958        &0.959        & 0.961 
\\ \hline
&& T=10K &&&\\ \hline
$a$ ($\AA$)     & 5.4812(1)   & 5.4845(1)   & 5.4823(1)   & 5.4858(1)   & 
5.4855(2)\\
$c$ ($\AA$)     & 13.2759(3)  & 13.2797(4)  & 13.2737(4)  & 13.2718(4)  
&13.2637(4) \\
$V$ ($\AA^{3}$) & 345.415(13) & 345.931(16) & 345.504(17) & 345.897(17) 
&345.642(17)\\
$m$ ($\mu_{B}$) & 3.445(24)   & 3.327(27)   & 3.160       & 2.272(50)   & 
0.727(93) \\
$\chi^{2}$ ($\%$) & 3.18      & 2.76        & 3.30        & 2.67        & 2.82 
\\
$W'$ 		  & 0.967     &0.968        &0.969        &0.966        &0.966 
\end{tabular}
\end{center}
\end{table}
 
\begin{figure}
\caption{Neutron diffraction patterns of
La$_{0.7}$Sr$_{0.3}$Mn$_{1-x}$Cu$_{x}$O$_{3}$(x=0.10) at room temperature and 
10 K.(The
bottom curves(Yobs-Ycal) are the difference between experimental
data and refinement data. The vertical bars indicate the
magnetic(bottom) and Bragg(top) peak positions). Arrows indicate magnetic 
diffraction peaks.}
\end{figure}

\begin{figure}
\caption {Average (Mn,Cu)-O bond lengths $d_{Mn-O}$(a), (Mn,Cu)-O-(Mn,Cu) bond 
angles $\theta_{\langle Mn-O-Mn\rangle}$(b) of 
La$_{0.7}$Sr$_{0.3}$Mn$_{1-x}$Cu$_{x}$O$_{3}$(x=0, 0.05, 0.10, 0.15, 0.20)  at 
room temperature and at 10 K.}
\end{figure}

\begin{figure}
\caption {The magnetization versus temperature(M-T) curves of 
La$_{0.7}$Sr$_{0.3}$Mn$_{1-x}$Cu$_{x}$O$_{3}$ (x=0.05, 0.10, 0.15) measured 
under field cooling (FC) and zero field cooling (ZFC) conditions in a magnetic 
field of 50 Oe.}
\end{figure}

\begin{figure}
\caption {Electric resistivity $\rho$ versus temperature for 
La$_{0.7}$Sr$_{0.3}$Mn$_{1-x}$Cu$_{x}$O$_{3}$(x=0.10(a), 0.15(b), 0.20(c)) in 
applied magnetic field H=0,1,3,5T.}
\end{figure}

\end{document}